\documentclass[12pt]{iopart}

\usepackage{graphicx}
\usepackage{iopams}  

\usepackage[breaklinks]{hyperref}

\begin{document}

\title[Resonance ionization spectroscopy of thorium isotopes]{Resonance ionization spectroscopy of thorium isotopes - towards a laser spectroscopic identification of the low-lying 7.6\,eV isomer of $^{229}$Th}

\author{S Raeder$^1$, V Sonnenschein$^{1,2}$, T Gottwald$^1$, I D Moore$^2$, M Reponen$^2$, S Rothe$^{1,3}$, N Trautmann$^4$, and K Wendt$^1$}

\address{$^1$ Institute of Physics, University of Mainz, Staudingerweg 7, 55128 Mainz, Germany}

\address{$^2$ Department of Physics, University of Jyv\"askyl\"a, Survontie 9, 40014 Jyv\"askyl\"a, Finland}

\address{$^3$ CERN, CH-1211 Genève 23, Switzerland}

\address{$^4$ Institute of Nuclear Chemistry, University of Mainz, Fritz-Stra\ss mann-Weg 2, 55128 Mainz, Germany}

\eads{\mailto{raeder@uni-mainz.de}, \mailto{volker.t.sonnenschein@jyu.fi}}

\begin{abstract}

In-source resonance ionization spectroscopy was used to identify an efficient and selective three step excitation/ionization scheme of thorium, suitable for titanium:sapphire (Ti:sa) lasers. The measurements were carried out in preparation of laser spectroscopic investigations for an identification of the low-lying $^{229\,m}$Th isomer predicted at $7.6\pm 0.5\,$eV above the nuclear ground state.
Using a sample of $^{232}$Th, a multitude of optical transitions leading to over 20 previously unknown intermediate states of even parity as well as numerous high-lying odd parity auto-ionizing states were identified. Level energies were determined with an accuracy of $~ 0.06\,\textrm{cm}^{-1}$ for intermediate and $0.15\,\textrm{cm}^{-1}$ for auto-ionizing states. Using different excitation pathways an assignment of total angular momenta for several energy levels was possible. One particularly efficient ionization scheme of thorium, exhibiting saturation in all three optical transitions, was studied in detail. For all three levels in this scheme, the isotope shifts of the isotopes $^{228}$Th, $^{229}$Th, and $^{230}$Th relative to $^{232}$Th were measured. An overall efficiency including ionization, transport and detection of $0.6\,\%\,$ was determined, which was predominantly limited by the transmission of the mass spectrometer ion optics.

\end{abstract}

\pacs{32.30.-r, 32.80.Rm, 32.80.Zb, 42.62.Fi}

\maketitle

\section{Introduction}
\label{Lbl:Introduction}

High resolution $\gamma$-spectroscopic experiments predict the existence of an isomeric state of $^{229}$Th, $^{229\, m}$Th,$\, I= {3 \over 2}^+$, with an excitation energy of only 7.6 $\pm$ 0.5\,eV \cite{Beck} above the ground state $^{229\, g}$Th,$\, I= {5 \over 2}^+$. This is the lowest excitation energy of an excited state in any nucleus known so far, lying in the range of typical atomic excitation and ionization energies. The $\gamma$-spectroscopic measurements used a differencing technique of $\gamma$-ray decay paths to infer the excitation energy from the observed spectra. Direct observation of the decay of the isomer via a M1 transition to the nuclear ground state was not successful. The reason for this was in part attributed to earlier expectations, which pointed towards a lower excitation energy of about 3.5\,eV \cite{Helmer}. This resulted in several searches for UV emission from $^{229\, m}$Th in an incorrect wavelength regime \cite{Irwin, Richardson, Shaw}. In addition, non-radiative decay paths, such as internal conversion and electron scattering \cite{Tkalyaa,Tkalyab}, as well as absorption of the 160\,nm VUV decay photons in most media might obscure the results and would require a dedicated experimental setup. The half-life of  $^{229\, m}$Th is theoretically estimated to be about 1 to 5 hours by extrapolation from a comparable transition of the $^{233\, m}$U isomer \cite{Beck} connecting the same Nilsson states but with an energy difference of 312\,keV, as well as from calculations based on another M1 transition in the $^{229}$Th nucleus \cite{Dykhne}. Nevertheless, large deviations from this estimate could be possible depending on the exact electronic configuration \cite{Tkalyaa}. 
Earlier calculations, based on the lower excitation energy of 3.5\,eV, neglected the possibility of internal conversion that becomes possible when accepting the energy of 7.6\,eV, which is higher than the ionization potential of 6.32\,eV for thorium. A recent experiment \cite{Kikunaga} hints at a half-life of less than 2\,hours, probably affected by the chemical conditions of the sample.\\
Aside from the basic nuclear physics interest, the low-lying isomer would provide a unique opportunity for fundamental research on laser-optical atom-to-nucleus coupling. For example, its use as a nuclear time standard of ultimate precision was proposed \cite{Peik}. 
With respect to state-of-the-art atomic clocks a nuclear clock could exhibit significant  advantages, e.g., the nucleus is shielded from its environment by the surrounding electron shell, preventing or at least tremendously reducing line-shifts induced by external fields. The currently estimated half-life of the isomeric state would result in a linewidth of less than 1\,mHz, leading to a theoretical precision of $\Delta\lambda / \lambda  \approx 10^{-19}$, outbalancing current high precision clocks by about two orders of magnitude. In addition, the strong dependence of the transition on the balance of individual fundamental forces within the nucleus could yield an amplification of several orders of magnitude of any frequency shift effect, e.g., induced by a potential variation of the electromagnetic coupling constant $\alpha$ with time \cite{Litvinova}, or of the constituent quark masses \cite{Flambaum}. In this manner, long term comparisons of a hypothetical $^{229\, m}$Th clock to a conventional atomic clock could stimulate the ongoing research on the cosmological evolution of the fundamental force coupling constants. Regarding the field of quantum optics and quantum computing, a nuclear isomeric state that could be directly interrogated by laser light also could serve as a dedicated qubit with extraordinary features. Finally, the isotope $^{229}$Th is  most likely a suitable candidate for NEET (Nuclear Excitation by Electron Transition), a process, which could help in pumping the isomeric state from the ground state \cite{Izosimov}.\\

In order to ascertain the existence of the predicted low-lying isomeric state of the nucleus $^{229}$Th, an alternative approach to gamma spectroscopy or other means of direct observation of the decay is needed. This could involve the identification of the isomer through a measurement of its specific hyperfine structure within a selected optical resonance in the atomic spectrum in comparison with to the well known structure related to the influence of the nuclear ground state coupling to the electron shell. In this work we describe laser spectroscopic investigations which serve as important preparation of such experimental activities.\\

One method to produce the radioisotope $^{229}$Th is via the alpha decay of the parent nucleus $^{233}$U, whereby a small branching ratio of about 2\,$\% $ of the decay populates the low-lying isomeric level \cite{Kikunaga}. In the last few years, efforts at the Jyv\"askyl\"a IGISOL iosotope separator facility, Finland, have concentrated on the development of a novel ion guide system optimized for the extraction of daughter recoil nuclei originating from the $\alpha$-decay of, e.g., a $^{233}$U source \cite{Tordoffb}. The source, electroplated onto stainless steel strips, is mounted on the inside surface of a helium-filled gas cell. Gas flow and an electrical guiding field efficiently extract the decay recoils as positively charged ions, primarily in a singly- or doubly-charged state. After a number of gas cell improvements the extraction efficiency for decay products of $^{233}$U, namely $^{221}$Fr (T$_{1/2}= 4.9$ min) and $^{217}$At (T$_{1/2}= 32$ ms), has been increased to about $16\,\%$. On the other hand, the extraction efficiency for $^{229}$Th does not exceed $1.6\,\%$ \cite{Moore08}, a factor of 10 lower. This fact is ascribed to an unusual and unexpectedly high fraction of produced thorium persisting in or converting into the neutral charge state. 
With a generation rate of $\approx 10^5$ recoils of $^{229}$Th per second in the present setup a $^{229\,m}$Th beam production with an intensity of approximately 30 ions$/$s is expected. For both the nuclear ground state as well as the isomeric atoms the optical excitation probability of each atomic transition is distributed over several hyperfine components differing in their total angular momentum. This fact significantly reduces optical line strength on a given wavelength of resonance. Combined with a limited fluorescence detection efficiency of typically only about $0.02\,\%$ and considerable background in a laser spectroscopy arrangement providing high resolution as, e.g., collinear fast beam laser spectroscopy \cite{Tordoffa}, successful unfolding of the isomeric hyperfine structure from that of the ground state is highly unlikely. However, considering a reasonably large fraction of neutral thorium recoils, the opportunity for increasing the extraction efficiency through the use of laser ionization via in-source resonance ionization on the neutral species, becomes highly favourable.\\

Laser Ion Sources (LIS) give access to efficient and selective ionization \cite{Geppert} at on-line facilities such as ISOLDE at CERN, Geneva, Switzerland, ISAC at TRIUMF, Vancouver, Canada or others \cite{Fedosseev08}. Today, state-of-the-art solid state laser systems are increasingly used for efficient multi-step resonant excitation and ionization of the atoms of a specific element. This technique of resonance ionization spectroscopy (RIS) is also under development for application at the IGISOL system in Jyv\"askyl\"a using Titanium-sapphire (Ti:sa) lasers. There it will be adapted for use in combination with the aforementioned gas cell and ion guide technique with a special focus on the production of $^{229}$Th ions. As a first step for preparation of a highly sensitive laser-based detection technique of the isomer $^{229\, m}$Th, Ti:sa laser RIS on thorium has been carried out in the LARISSA laboratory at Mainz University, Germany, in order to identify suitable resonant laser excitation and ionization schemes. \\

A number of laser spectroscopic investigations on thorium have been performed earlier, e.g., for analytics or for addressing the study of isotope shifts and the hyperfine structure of $^{229}$Th in order to determine nuclear charge radii and nuclear moments \cite{KALBER89}. Previous RIS investigations on thorium applied pulsed as well as continuous-wave dye lasers within a simple two-step excitation/ionization scheme. The pulsed laser RIS localized low-lying excited levels of odd parity around $26\,000\, \textrm{cm}^{-1}$ as well as numerous even parity auto-ionizing levels just above the ionization potential at $50\,867\, \textrm{cm}^{-1}$ \cite{Johnson93}. Due to the use of high laser power considerable background was generated, which hampered the foreseen development of a high precision, high accuracy determination technique for thorium isotope ratios in geological samples \cite{Billen93}. Correspondingly, the technique of continuous wave laser RIS was also established by one of the groups using non-resonant ionization by a high power argon ion laser. An impressive overall efficiency of this technique of up to $ 0.4$\,\% has been demonstrated \cite{Johnson93}. In the present study an excitation/ionization scheme involving three resonant steps is favoured, which promises comparable or even higher efficiency in combination with high suppression of background from uranium, which could be generated either by non-resonant ionization or by fragmentation of uranium oxide induced by any blue to UV laser radiation.

\section{Experimental setup}

\begin{figure}
\centering
\includegraphics[width=0.6\textwidth]{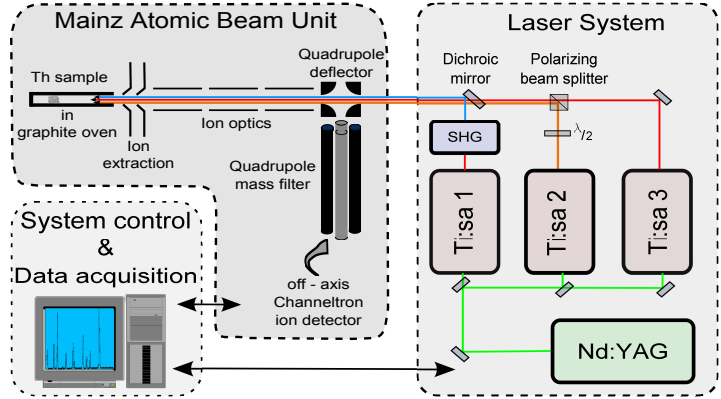}

\caption{\label{LBL:Fig:Aufbau}Schematic experimental setup: laser system consisting of three Ti:sa lasers and a Nd:YAG pump laser, mass spectrometer with graphite oven, quadrupole mass filter and ion detector. The components are controlled by the data acquisition system. }
\end{figure}

The laser system used for the measurements presented in this work is composed of a set of three tunable nanosecond Ti:sa lasers jointly pumped by a commercial frequency doubled Nd:YAG laser (Photonics Industries DM 532-80) with a repetition rate of typically 10\,kHz. Output power reaching up to 3\,W is obtained with a pump laser power of about 16\,W per Ti:sa laser. Frequency selection is achieved by a combination of birefringent filter and etalon within the laser resonator \cite{Mattolat}, which results in a spectral bandwidth of $~5$\,GHz. However, such a laser construction provides only limited wavelength tunability. Therefore, one of the lasers is equipped with a grating mounted in Littrow configuration for frequency selection \cite{Haensch}. This laser exhibits a continuous frequency tuning range from 710\,nm up to 950\,nm, which is ideally suited for the long range search of unknown atomic transitions. Single pass second harmonic generation (SHG) of the fundamental Ti:sa laser radiation in an optical nonlinear crystal ($\beta$-BaB$_2$O$_4$) is employed to obtain up to 500\,mW output power in the blue to UV spectral range. The typical pulse length is approximately 40\,ns. Pulses of individual lasers are temporally synchronized within $5\,$ns using fast Pockels cell switching. The time structure as well as the wavelength of each laser is monitored by fast photodiodes and commercial wavemeters (High Finesse WS6, ATOS LM007), respectively. The laser beams are finally overlapped and focussed, resulting in a beam waist of about 100\,$\mu$m within the center of the graphite oven (inner diameter 2.2\,mm, 50\,mm length) of the MABU (Mainz Atomic Beam Unit) spectroscopic setup. A schematic sketch of the latter is given in Figure \ref{LBL:Fig:Aufbau}.  %
\par
Samples for spectroscopic investigations contained about $10^{15}\,$ atoms of $^{232}$Th in nitric acid solution. Each solution was dried on a small piece of zirconium foil, which was folded for improved reduction of thorium oxides into neutral thorium, and inserted into a graphite oven. A long-term stable atom source was achieved by carefully heating the oven resistively up to 1600 K. The evaporated thorium atoms are then resonantly excited and subsequently ionized by laser radiation inside the oven. The resulting photo ions, as well as thermally produced interfering surface ions, were extracted and guided by ion optics to a static quadrupole deflector. Deflecting the ion beam by $90^\circ$ separated the ions from the neutral particles and allowed for a convenient anti-collinear overlap geometry of laser radiation and atomic beam. A subsequent quadrupole mass filter was used to suppress surface ions of different masses produced in the hot cavity. To ensure reasonable transmission in the order of $10\, \%$ the mass filter was tuned to a somewhat reduced mass resolution of $m/\Delta m \approx 200$ during the spectroscopic experiments. Ions of interest were finally detected quantitatively by an off-axis channeltron detector operated in single ion counting mode. Data acquisition was performed with a LabView program and included reading of the ion count rate, monitoring of the wavelengths of the lasers run with fixed frequencies as well as the control of the grating laser tuning and the quadrupole mass filter settings. 

\section{Measurements and Results}

An efficient excitation and ionization scheme for atomic thorium requires knowledge of suitable high-lying energy levels as well as auto-ionizing states above the ionization potential (IP). Existing information on the spectrum of thorium is quite limited, with transitions and configurations known only for energy levels up to $38\,000\,\textrm{cm}^{-1}$ and the ionization potential of thorium is reported to be $50\,867 (2)\,\textrm{cm}^{-1}$ \cite{Kohler}. Correspondingly, for all investigations reported in this study three step ionization schemes with the first excitation step in the blue to UV wavelength region followed by two excitation steps in the infrared, were selected.

\subsection{Intermediate states}

\begin{figure}
\centering
\includegraphics[width=0.75\textwidth]{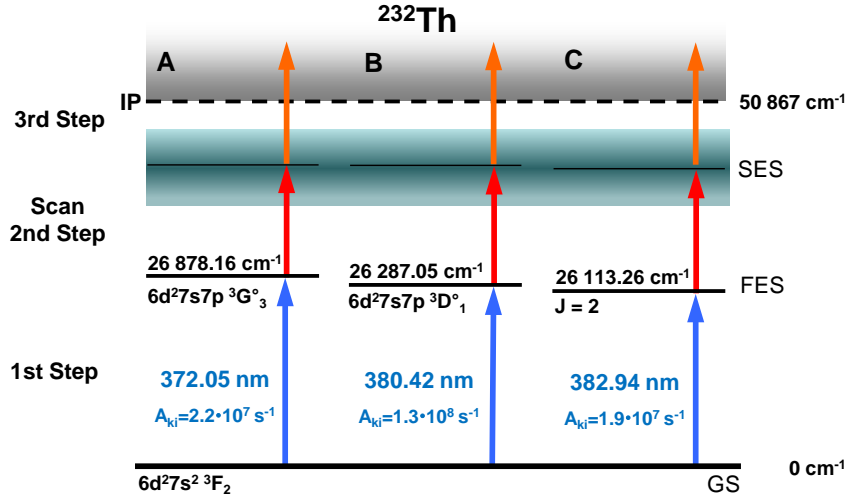}
\caption{\label{LBL:Fig:secsteps}Selected strong transitions for the first excitation step of suitable ionization schemes for thorium. $J$-Values, $A$-factors and configurations are taken from \cite{Kurucz,Blaise}.}
\end{figure}%

\begin{figure}
\centering
\includegraphics[width=0.99\textwidth]{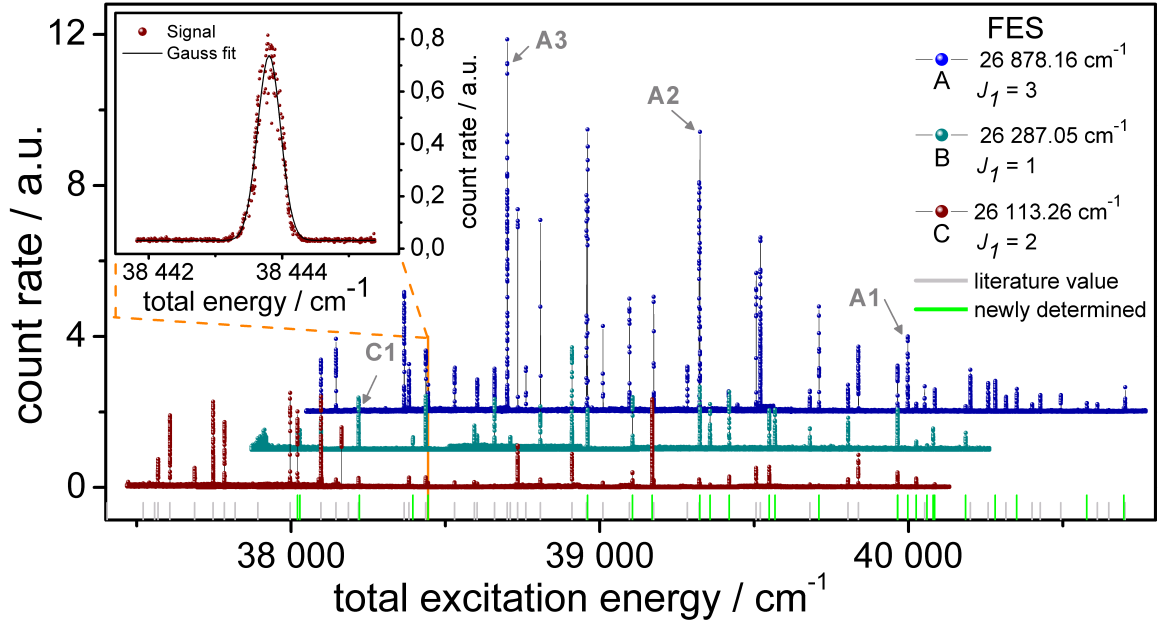}
\caption{\label{LBL:Fig:Secspectra} Three scans for high lying levels of even parity starting from different FES of odd parity, denoted A, B and C according to Figure 2. For clarity the individual scans are separated by an artificial offset. Resonances addressed in further studies are indicated by A1, A2, A3 and C1. The inset shows a detailed scan of one of the intermediate states used for the precise evaluation of level energies. At the bottom line earlier known and newly determined resonances are marked by gray and red dashes, respectively.}
\end{figure}

Three well known energy levels \cite{Kurucz,Blaise} accessible by frequency doubled Ti:sa laser radiation and exhibiting high transition strengths from the atomic ground state (GS) were chosen as first excited states (FES) as shown in Figure \ref{LBL:Fig:secsteps}. For further excitation into unknown high-lying excited states from one of these FESs, the grating-assisted Ti:sa laser was scanned over an extended wavelength range from 715\,nm to 880\,nm with a spectral step width of about 0.01\,nm (0.3\,GHz). A third laser operating at a wavelength around 730\,nm ensured non-resonant photo-ionization from levels populated by the scanning laser. Numerous even parity levels in an energy region from  $37\,500\, \textrm{cm}^{-1}$ to  $40\,500 \, \textrm{cm}^{-1}$ with a total angular momentum $J_2=J_1-1, J_1, J_1+1$, where $J_1$ indicates the angular momentum of the FES, were identified. Information on the angular momentum of each energy level was obtained by exciting the same energy level from different FES with different $J$-values. Figure\,\,\ref{LBL:Fig:Secspectra} shows the combined laser scans from all three FESs with an artificial individual offset introduced for visual clarity. In these spectra, 24 transitions to previously unknown levels were identified in addition to the confirmation of 43 levels already mentioned in the literature \cite{Blaise}. Finally, detailed laser scans with high spectral resolution were performed for all newly observed levels as well as for a few of the earlier known levels for reference. %

\begin{table}
\caption{\label{TBL:secstates} Newly identified high-lying energy levels of even parity in atomic thorium. Signal strength indicates the intensity of the ion signal on each resonance populated, as related to one of the three FESs, denoted A, B, C according to Figure \ref{LBL:Fig:secsteps}. Labeling is: strong (S), medium (M), weak (W) or not visible (n). The excitation of the individual levels from different first excited steps constrain the $J$-values; possible values are given. In case of ambiguity the most probable $J$-value is printed in bold.} 

\begin{indented}
\lineup
\item[]\begin{tabular}{cccc}
    \br 
    & {\bf Energy}& {\bf \scriptsize Signal strength} & {\bf $J$}\\
    &  cm$^{-1}$& {\scriptsize from levels A,B,C} & \\
    \mr
    1     & $38\,021.14(6)$ & n,W,S & {\bf 1},2 \\
    2     & $38\, 028.82(8)$ & n,W,n & {\bf 0},1,2 \\
    3     & $38\, 221.68(8)$ & n,W,n & {\bf 0},1,2 \\
    4     & $38\, 394.92(8)$ & n,M,n & {\bf 0},1,2 \\
    5     & $38\, 443.82(6)$ & M,n,W & {\bf 3},2 \\
    6     & $38\, 960.68(7)$ & S,S,W & 2 \\
    7     & $39\, 106.15(5)$ & n,S,M & {\bf 1},2 \\
    8     & $39\, 324.25(7)$ & S,S,W & 2 \\
    9     & $39\, 357.28(4)$ & n,S,W & {\bf 1},2 \\
    10    & $39\, 419.39(8)$ & n,S,M & {\bf 1},2 \\
    11    & $39\, 548.81(5)$ & n,S,M & {\bf 1},2 \\
    12    & $39\, 567.43(8)$ & n,S,n & {\bf 0},1,2 \\
    13    & $39\, 710.05(6)$ & S,n,W & 2,{\bf 3} \\
    14    & $39\, 964.65(6)$ & M,S,M & 2 \\
    15    & $39\, 997.59(8)$ & M,n,n & 2,3,{\bf 4} \\
    16    & $40\, 024.98(8)$ & W,M,M & 2 \\
    17    & $40\, 059.75(8)$ & n,W,n & {\bf 0},1,2 \\
    18    & $40\, 079.71(5)$ & n,M,W & {\bf 1},2 \\
    19    & $40\, 084.72(8)$ & W,n,W & 2,{\bf 3} \\
    20    & $40\, 185.14(5)$ & W,M,S & 2 \\
    21    & $40\, 280.93(8)$ & M,-,- & 2,3,4 \\
    22    & $40\, 350.52(8)$ & M,-,- & 2,3,4 \\
    23    & $40\, 577.38(8)$ & W,-,- & 2,3,4 \\
    24    & $40\, 697.45(8)$ & W,-,- & 2,3,4 \\    
 \br   
    
\end{tabular}
\end{indented}
\end{table}%
The experimental data of each individual resonance were found to be well described by a saturated Gaussian function given by
\begin{eqnarray}
y = y_0 + A_0 \cdot \frac{S_0 \cdot G(E_0,w_G)}{S_0 \cdot G(E_0,w_G)+1} \label{Eqn:saturation} \\
\textrm{with}\nonumber\\
G(E_0,w_G)= \frac{1}{ \sqrt{2 \pi \cdot w_G^2}} \cdot \textrm{e}^\frac{-(E-E_0)^2}{2 \cdot w_G^2}
\end{eqnarray}
and the centroid energy $E_0$, the Gaussian linewidth $w_G$, the saturation parameter $S_0$, the amplitude of the peak $A_0$ and an offset $y_0$. This confirms the assumption that the spectroscopic signal is dominated by the Gaussian spectral distribution of the laser convoluted with similarly shaped Doppler profile of the atom ensemble inside the source, while a contribution from the natural linewidth is not visible. An example for such a fit is shown in the inset of Figure \ref{LBL:Fig:Secspectra}, exhibiting a FWHM of about $0.2\,\textrm{cm}^{-1}$. 
All newly observed levels of even parity as discovered in this work are tabulated in Table\,\ref{TBL:secstates} with their centroid energy, relative signal strength and possible $J$-values. The latter are restricted by the excitation of the individual levels from different FESs; with the most probable $J$-value printed in bold in case of ambiguity. The statistical error for the centroid energy, as determined from several measurements of the same level, is given. The relative signal strength, as indicated only qualitatively in Table\,\ref{TBL:secstates} by a letter code, is specific for the experimental arrangement and the conditions used in this work. However, this rough classification is sufficient for the identification of an efficient excitation scheme, while absolute transition strengths could not be determined with the present setup.

\subsection{Auto-ionizing and Rydberg states}

\begin{figure}
\centering
\includegraphics[width=0.75\textwidth]{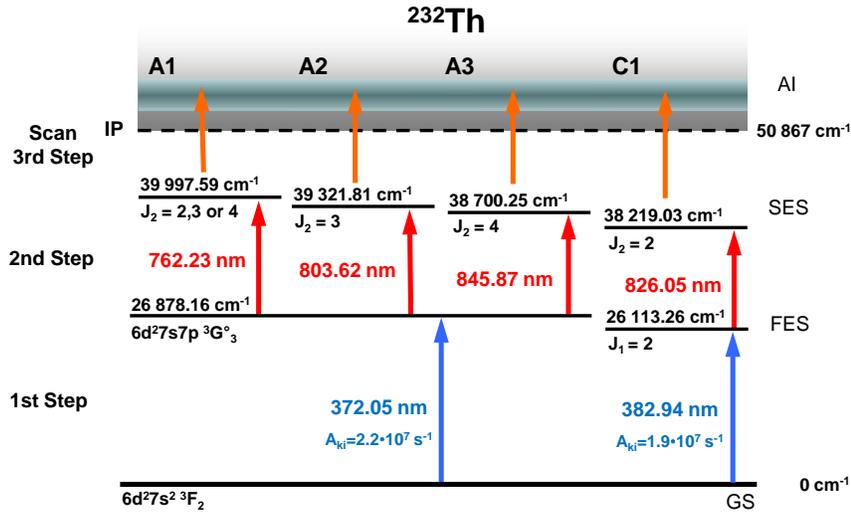}

\caption{\label{LBL:Fig:Aisteps}Schemes for the search for auto-ionizing resonances of odd parity. From four SESs of even parity and different $J$-value, laser scans for excitation just above the ionization potential were performed.}
\end{figure}

\begin{figure}
\centering
\includegraphics[width=0.99\textwidth]{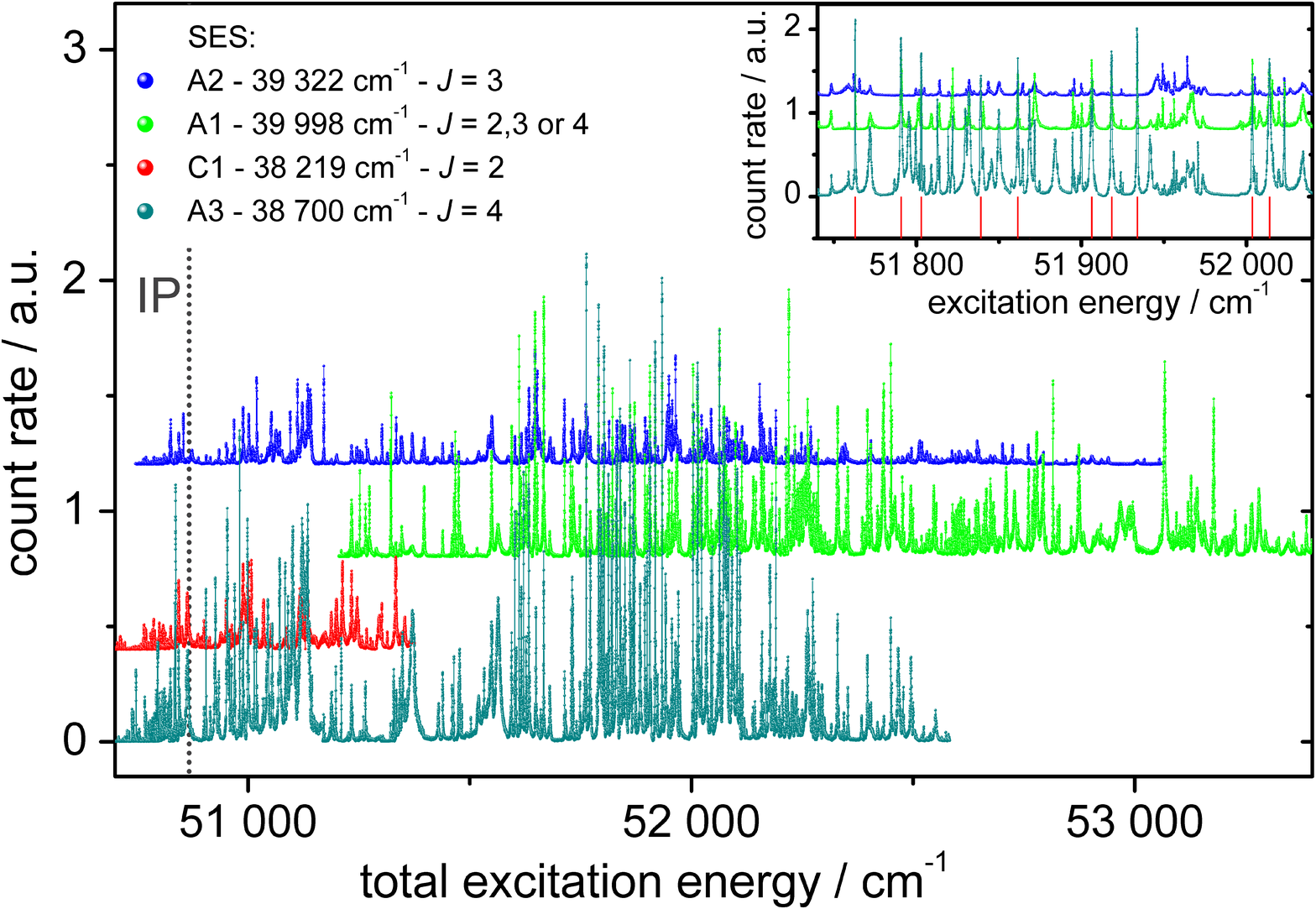}

\caption{\label{LBL:Fig:Aispectra}Spectra of AI resonances just above the IP, which is indicated at the left hand side. A very high level density is observed. Excitation started from one of four different high-lying SESs as given. The inset shows the region of $51\,700\, -\, 52\,100\,\textrm{cm}^{-1}$ with the strongest AI resonances in more detail. The energetic position of resonances used for further investigation as summarized in Table\,\ref{LBL:AIsteps} are indicated by green dashes.}
\end{figure}

Suitable final excitation steps towards ionization were determined in a similar manner as the second excited states (SES). While the first two excitation steps were kept fixed on selected resonances, the grating-tuned Ti:sa laser was scanned to cover the region ranging from $50\,700\, \textrm{cm}^{-1}$ to  $53\,000 \, \textrm{cm}^{-1}$ of total excitation energy. The four intermediate excitation steps with strong ion signal, which were chosen for these further investigations, are shown in Figure \ref{LBL:Fig:Aisteps}. As visualized in Figure \ref{LBL:Fig:Aispectra} the level density observed in the energy region covered is much higher than the one in the region of the SESs around $38\,000\,\textrm{cm}^{-1}$, which is shown in Figure \ref{LBL:Fig:Secspectra}. All spectra recorded are highly complex, containing numerous peaks with different linewidths. In addition, overlapping of different structures and the occurrence of significant interference effects are observed. Correspondingly, no Rydberg series could be identified unambiguously in any of the spectra.

\begin{table}
\caption{\label{LBL:AIsteps} Compilation of the ten strongest AI resonances of odd parity investigated. All resonances were observed starting from the same SES at $38\,700.25\,\textrm{cm}^{-1}$ from scheme $A3$. The relative line strength indicates the ion signal relative to the strongest AI level, which is denoted level\,1. A restriction on possible $J$-values is deduced from comparing excitation from SESs with different $J_2$.} 

\begin{indented}
\lineup
\item[]\begin{tabular}{ccccc}
    \br
    {\bf Level} & {\bf Energy} & {\bf rel. strength} & {\bf width} & {\bf $J$} \\
          & cm$^{-1}$  &       & cm$^{-1}$  &  \\
    \mr
    1     & $51\, 762.84(15)$ & 1.0   & 0.25  & 3,4,5 \\
    2     & $51\, 790.65(15)$ & 0.8   & 0.60  & 3,4 \\
    3     & $51\, 802.84(15)$ & 0.8   & 0.26  & 3,4,5 \\
    4     & $51\,839.03(15)$ & 0.7   & 0.46  & 3,4,5 \\
    5     & $51\, 861.38(15)$ & 0.8   & 0.34  & 3,4 \\
    6     & $51\, 906.20(15)$ & 0.7   & 0.73  & 3,4 \\
    7     & $51\, 918.27(15)$ & 0.9   & 0.45  & 3,4 \\
    8     & $51\, 933.81(15)$ & 1.0   & 0.36  & 3,4,5 \\
		9     & $52\,003.55(15)$ & 0.7   & 0.34  & 3,4,5 \\
    10    & $52\, 014.01(15)$ & 0.7   & 0.90  & 3,4,5 \\   
  \br
    \end{tabular}
\end{indented}
\end{table}

The inset in Figure \ref{LBL:Fig:Aispectra} shows a detailed view of the most interesting region for the determination of a suitable, highly efficient ionization scheme. It enlarges the region of strongest transition strengths but also of highest level density. A number of narrow transitions with resonance widths comparable to the laser linewidth of 5\,GHz is observed here. These auto-ionizing levels provide very high ionization rates, a factor of $\sim$100 above the background of non-resonant ionization. \Tref{LBL:AIsteps} indicates the ten strongest auto-ionizing resonances determined in these scans, all starting from the same SES $(A3)$ located at $38\,700.25 \, \textrm{cm}^{-1}$ and exhibiting rather similar ionization probability.

\subsection{Characterization of the ionization scheme}

\subsubsection{Saturation}

\begin{figure}
\centering
\includegraphics[width=0.28\textwidth]{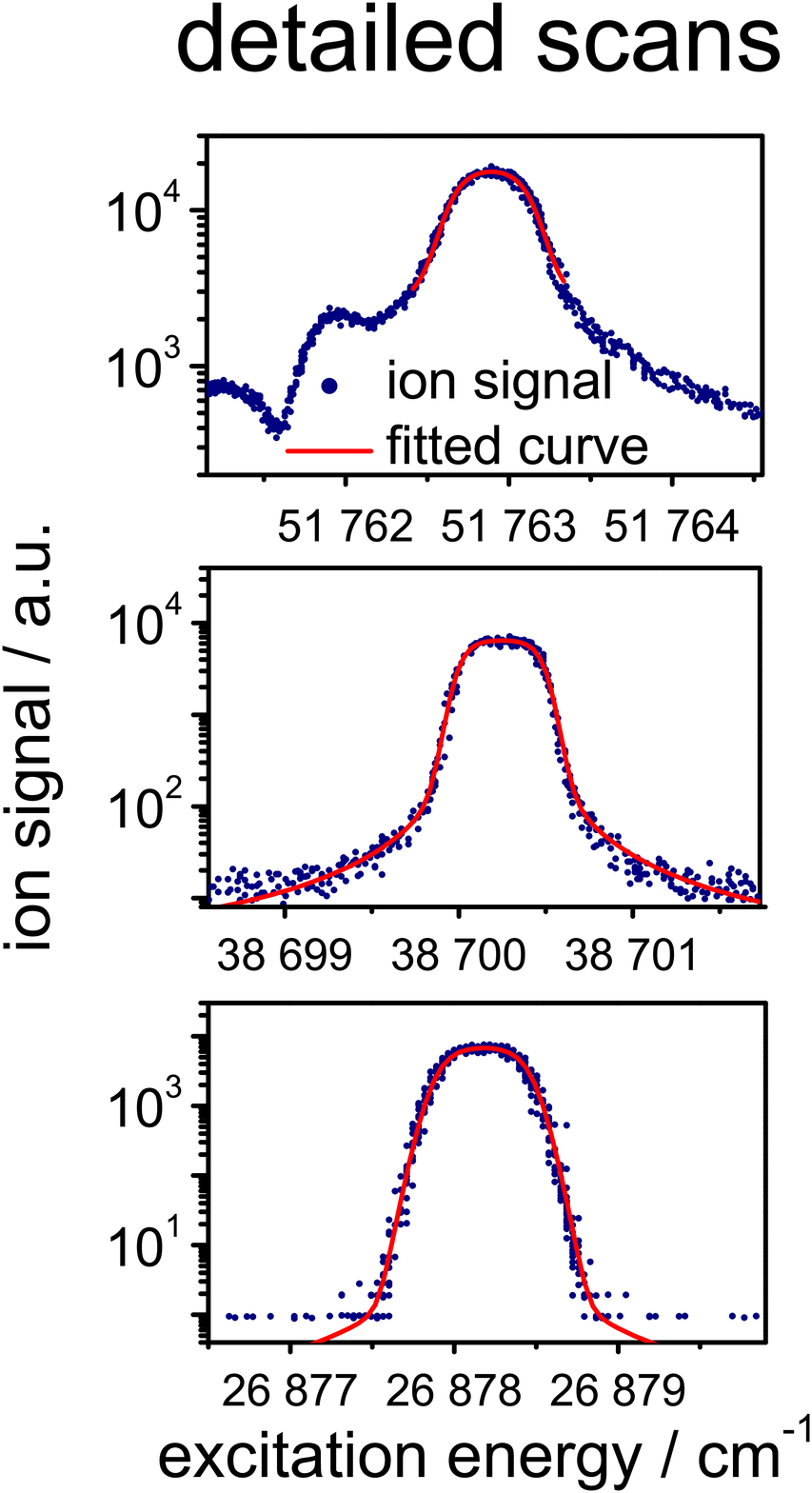}
\hspace{0.2cm}
\includegraphics[width=0.37\textwidth]{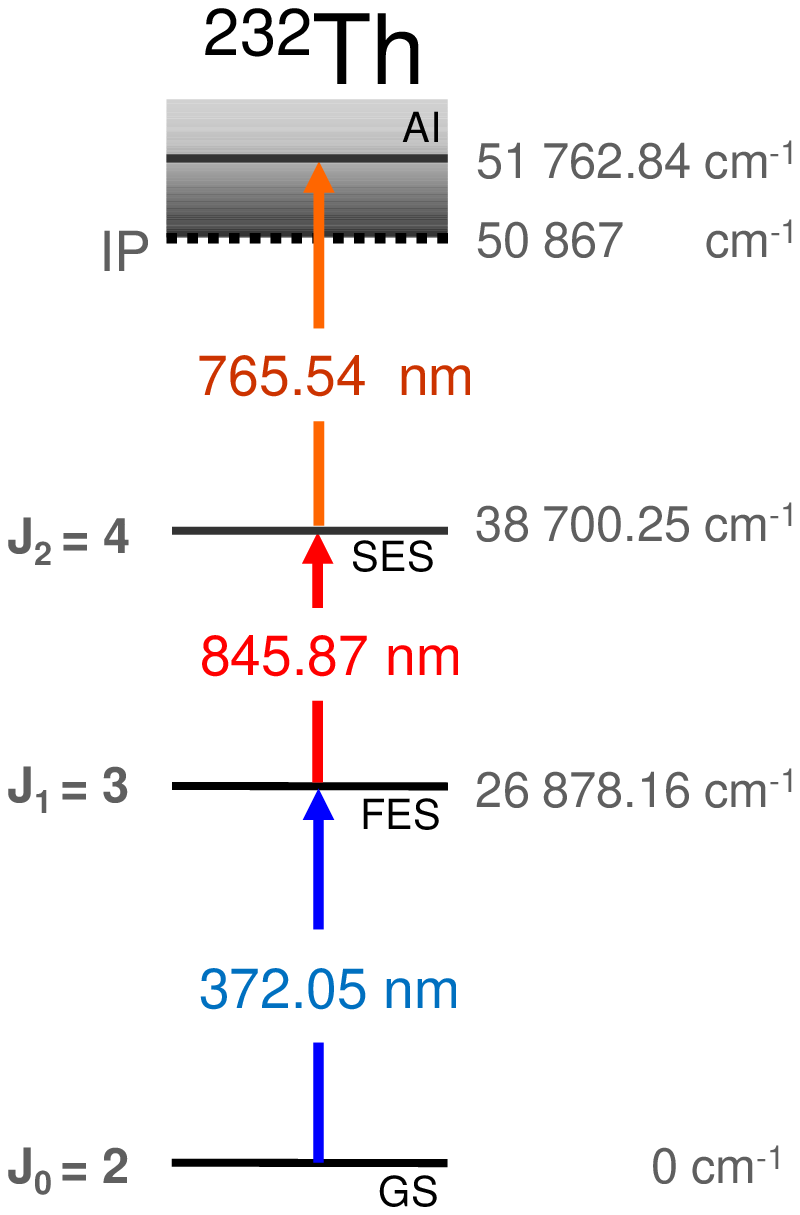}
\hspace{0.2cm}
\includegraphics[width=0.24\textwidth]{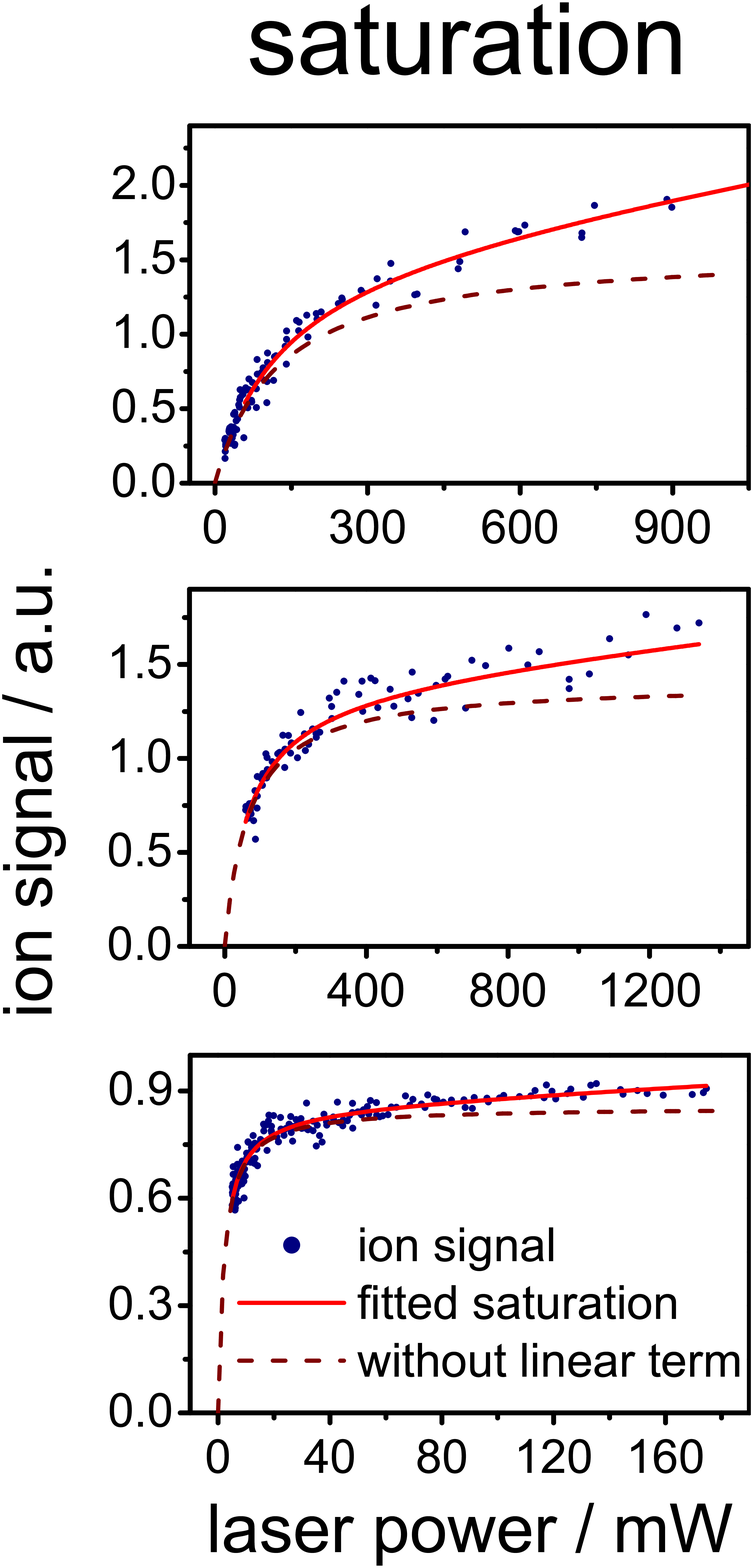}
\caption{Final three step ionization scheme for $^{232}$Th with detailed scans for each excitation step (left) and saturation curves (right).\label{LBL:Fig:finalscheme}}
\end{figure}

The excitation scheme finally chosen involves a strong, yet narrow auto-ionizing level at $51\,762.84\, \textrm{cm}^{-1}$. The scheme is depicted in Figure \ref{LBL:Fig:finalscheme} in combination with the saturation behavior as well as detailed spectral scans of each step. The latter were performed using significantly higher laser power for better statistics and correspondingly a slight Lorentzian contribution in the line shape appeared due to saturation effects. Peaks were thus fitted with a saturated Voigt profile, in which the Voigt convolution was used in \ref{Eqn:saturation} instead of the simple Gaussian distribution. The third excitation step towards ionization shows an interference structure together with a weak overlapping peak. Both effects are clearly visible in the logarithmic scaling chosen but cannot be assigned unambiguously to any known configuration.

For saturation measurements the ion signals were recorded as function of the laser power of each one of the excitation steps, while the power in the remaining two excitation steps were kept fixed. The resulting saturation curves were fitted with a function of the form
\begin{equation} 
f(P) = I_0 + m \cdot P + A \cdot\frac{P/P_S}{(1+P/P_S)}, 
\end{equation}
which extends the conventional saturation formula for a two-level atomic excitation with an additional linear term. The offset parameter $I_0$ accounts for the thermal ionization as well as non-resonant photo ionization, while $P$ represents the laser power and $P_S$ the saturation power. The linear increase with a slope $m$ accounts for deviations from the basic saturation behavior, which arise from non-resonant ionization processes driven by the laser whose power is varied during the measurement.  

As shown in Figure\,\ref{LBL:Fig:finalscheme} by the dotted curve, which omits the linear non-resonant ionization term, all transitions in the excitation scheme of interest can be fully saturated with the available laser powers. Corresponding saturation power values measured in this setup are 2.2\,mW for the first transition, 68\,mW for the second and 125\,mW for the third transition, respectively. Utilizing the totally available laser power for each excitation step, it can be estimated that saturation could be obtained on an atom-laser interaction area about ten times larger than the $0.03\, \textrm{mm}^{2}$ used here. This fact could prove useful in future experiments, for example regarding the rather large gas cell volume of the $^{229}$Th recoil ion guide mentioned previously.

\subsubsection{Isotope shifts and hyperfine structure}

\begin{figure}
\centering
\includegraphics[width=0.99\textwidth]{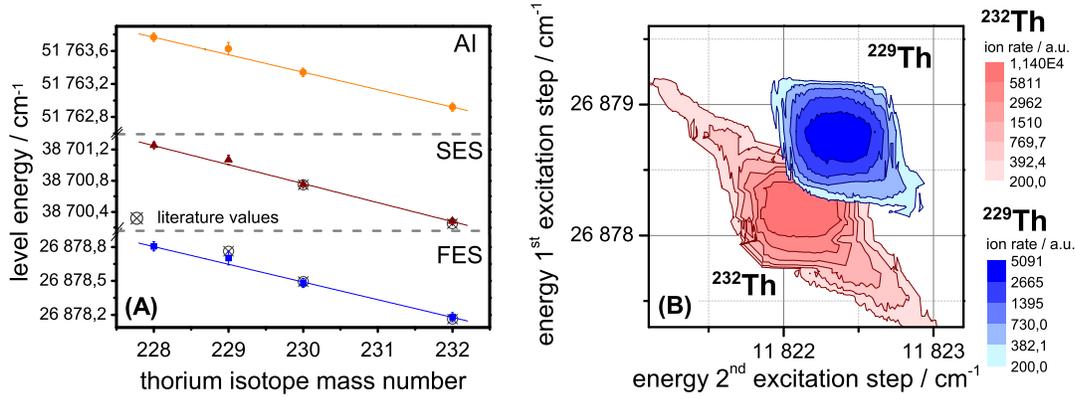}
\caption{$(A)$: Isotope shift of excited states in the ionization scheme of Figure\,\ref{LBL:Fig:finalscheme} for $^{228}$Th, $^{229}$Th, $^{230}$Th, and $^{232}$Th. Literature values are given for $^{230}$Th/$^{232}$Th, taken from \cite{Blaise}, and for $^{229}$Th, taken from \cite{Vernyi}. $(B)$: Two dimensional scan of the first and second excitation step for $^{229}$Th versus $^{232}$Th for the first two excitation steps.
\label{LBL:Fig:Isotopeshift}}
\end{figure}

\begin{table}
\caption{\label{Tab:Isotopeshift} Energy level positions of the ionization scheme for the isotopes $^{228}$Th,$^{229}$Th,$^{230}$Th, and $^{232}$Th and level isotope shifts; literature values as indicated.} 

\begin{indented}
\lineup
\item[]\begin{tabular}{l@{ }r*{4}{l}}

    \br
          &       & {\bf $^{228}$Th} & {\bf $^{229}$Th}  & {\bf $^{230}$Th}  & {\bf $^{232}$Th}  \\
    \mr
    \multicolumn{5}{l}{\bf first excited state (FES)}      &  \\
    measured & $\textrm{cm}^{-1}$  & $26\,878.81(4)$ & $26\,878.70(6)$ & $26\,878.48(4)$ & $26\,878.18(4)$ \\
    literature & $\textrm{cm}^{-1}$  &       & $26\,878.76^{\rm c}$ & $26\,878.49^{\rm b}$ & $26\,878.16^{\rm b}$ \\
    Level IS$^{\rm a}$ & $\textrm{cm}^{-1} $& $0.63(4)$ & $0.52(6)$ & $0.30(4)$ & $0$ \\
          &       &       &       &       &  \\
    \multicolumn{5}{l}{\bf second excited state (SES)}       &  \\
    measured & $\textrm{cm}^{-1}$  & $38\,701.25(4)$ & $38\,701.07(8)$ & $38\,700.76(4)$ & $38\,700.28(4)$ \\
    literature & $\textrm{cm}^{-1}$  &       &       & $38\,700.75^{\rm b}$ & $38\,700.25^{\rm b}$ \\
    Level IS$^{\rm a}$& $\textrm{cm}^{-1} $  & $0.97(4)$ & $0.79(6)$ & $0.47(4)$ & $0$ \\
          &       &       &       &       &  \\
    \multicolumn{5}{l}{\bf auto-ionizing state (AI)}        &  \\
    measured & $\textrm{cm}^{-1}$  & $51\,763.77(7)$ & $51\,763,63(9)$ & $51\,763,34(5)$ & $51\,762,92(5)$ \\
    Level IS$^{\rm a}$ & $\textrm{cm}^{-1} $ & $0.85(4)$ & $0.71(6)$ & $0.42(4)$ & $0$ \\    
		\br
    \end{tabular}
    \item[] $^{\rm a}$ Isotope Shift ($^{A}$Th - $^{232}$Th),\,\, $^{\rm b}$ taken from \cite{Blaise}, \,\, $^{\rm c}$ taken from \cite{Vernyi}.
  \end{indented}
\end{table}

The determination of the isotope shifts as well as the prediction of a potential isomer shift for the individual energy levels of the atomic excitation scheme is of utmost importance. Clearly, for investigations on the low-lying isomer in $^{229}$Th, the most interesting shift is the one between the $^{229}$Th nuclear ground state and the isomeric state. The thorium nitrate sample used in this work also contained minor amounts of $^{228}$Th and $^{230}$Th. Thus, for completeness, the corresponding isotope shifts for these isotopes relative to $^{232}$Th were determined. All data are given in Table\,\ref{Tab:Isotopeshift} as absolute level energies for each isotope together with level isotope shifts. Transition isotope shifts may be deduced by subtraction. Isotope shifts of $^{230}$Th for the first and second excited levels were reported already earlier in literature \cite{Blaise}, while for $^{229}$Th only the isotope shift for the first step was quoted in \cite{Vernyi}. No earlier data exists for the isotope $^{228}$Th. All values measured in this work agree very well with the previous results, as illustrated in Table\,\ref{Tab:Isotopeshift}. Figure \ref{LBL:Fig:Isotopeshift} $(A)$ graphically visualizes the level isotope shifts for each individual excitation, plotted as total level energies. The plot indicates a linear trend for the even A isotopes together with a weak odd-even staggering for the isotope $^{229}$Th as expected. Due to a different saturation behavior in the unresolved hyperfine components of the odd $A$ isotope $^{229}$Th, the energy level positions determined in this work from the unresolved peak structure do not necessarily represent the center of gravity of the hyperfine splitting precisely. The size of this effect can be estimated from the widths and the very slight asymmetries observed in the individual resonance lines. Errors on the energy level positions for the isotope $^{229}$Th were enlarged correspondingly. In general, the relative isotope shifts for each individual energy level correspond well to the changes in mean-square nuclear charge radii of thorium isotopes as given by \cite{KALBER89}. This behavior is expected in this high $Z-$region where the volume effect dominates the isotope shift and a possible mass effect can be neglected.

A two dimensional spectrum of the isotope shifts of $^{229}$Th versus $^{232}$Th in the first two levels is given in Figure \ref{LBL:Fig:Isotopeshift},$(B)$, depicting a limited optical isotope selectivity with slight overlap of the two peaks. The two dimensional resonance peak shapes and widths are dominated by the laser line width and the scanning step size. The pronounced diagonal peak enhancement predominantly visible for the much stronger signal of $^{232}$Th is induced by a resonance enhanced two photon process with both excitation steps being near-resonant and the sum energy exactly matching the level energy of the SES. As briefly mentioned above, no indication of the $^{229}$Th hyperfine structure has been observed in either one of the two transitions, indicating that its overall spread is always well below the experimental linewidth of $0.3\,\textrm{cm}^{-1}$.

\subsubsection{Ionization efficiency}

\begin{figure}
\centering
\includegraphics[width=0.5\textwidth]{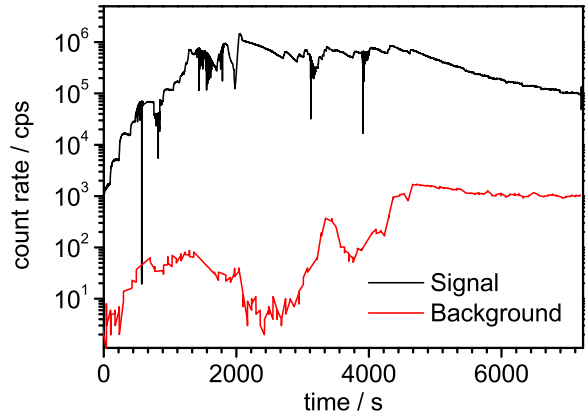}
\caption{Efficiency measurement on a sample of $4 \times 10^{11}$ atoms of $^{229}$Th. The furnace temperature was stepwise increased and the signal of laser-ions as well as of non-resonant background was recorded as a function of time.
\label{LBL:Fig:effmeass}}
\end{figure}

The overall ionization efficiency was determined by evaporating a precisely known number of atoms of $^{229}$Th until complete exhaustion and accumulation of the corresponding resonance ion counts in the detector, as shown in Figure\,\ref{LBL:Fig:effmeass}. The signal showed highest release of thorium at about 1900\,K oven temperature, while maximum heating reached up to 2300\,K. From the behavior of the count rate as function of the steadily increased oven temperature we conclude that at this maximum value the sample is almost totally evaporated. An efficiency value of $0.6\,\%$ was obtained by substracting the background from the integrated signal. The value was reproduced with only marginal variations in three individual measurements. It includes all possible losses within the individual processes of evaporation, chemical reduction of the oxide, atomization, resonance excitation and ionization and ion beam transmission through the apparatus. In the present setup, which was optimized for spectroscopic application instead of analytics, the latter is expected to cause the dominant limitation. From ion optical simulations the transmission is estimated to be of the order of only about $10\,\%$. Thus a further significant enhancement of the overall efficiency for in-source resonance ionization of Th is expected by correspondingly optimizing the transmission with the trade-off of a reduced mass resolution or by alternatively chosing a high transmission magnetic sector field mass separator for these experiments.

\section{Conclusion and Outlook}

For the identification of a suitable efficient three-step resonance ionization scheme for thorium, in-source resonance ionization spectroscopy was carried out. From hundreds of observed resonances the level energies and restrictions on the angular momenta for 24 previously unknown intermediate atomic energy levels and ten strong auto-ionizing states were extracted. A dedicated resonance ionization scheme was selected and characterized by measurements on saturation power and isotope shifts, while the hyperfine structure of the odd-$A$ isotope $^{229}$Th could not be resolved. An overall ionization efficiency of $0.6\,\%$ was determined for our experimental arrangement, which agrees well with the expectations. This excitation scheme will be used in the near future at the $^{229}$Th source for experiments at the Jyv\"askyl\"a IGISOL facility. By accessing the large neutral fraction of the recoil nuclei a significant efficiency increase in the low-energy radioactive ion beam production is envisaged. In this manner, the required beam intensity for high resolution collinear fast beam laser fluorescence spectroscopy on the hyperfine structure of the nuclear ground and isomeric state of $^{229}$Th may hopefully become possible.

A high resolution measurement of the hyperfine structure of the ground state of $^{229}$Th has recently been realized, utilizing an injection-locked narrow bandwidth Ti:sa laser. These data are presently under evaluation and the results will be published soon. In order to finally fully establish the existence of the still mysterious $^{229m}$Th isomer via its atomic hyperfine structure, these kind of measurements serve as an important prerequisite by delivering a precise template of the nuclear ground state and the corresponding hyperfine pattern beforehand. 

\ack 

This research was funded by the "Bundemisministerium f\"ur Bildung und Forschung" under contract number BMBF-06Mz228 and supported by the GRASPANP Graduate School in Particle and Nuclear Physics of Finland.


\section*{References}
\bibliographystyle{unsrt}
\bibliography{Thorium2}

\end{document}